\documentclass[3p,10pt,a4paper,twoside,fleqn,sort&compress]{filomat}
\usepackage{amsmath,latexsym}
\usepackage{amsthm}
\usepackage[varg]{pxfonts}
\usepackage{verbatim}

\newcommand{\FF}{\mathcal{F}}

\newcommand{\FilVol}{xx}
\newcommand{\FilYear}{yyyy}
\newcommand{\FilPages}{zzz--zzz}
\newcommand{\FilDOI}{(will be added later)}
\newcommand{\FilReceived}{Received: dd Month yyyy; Accepted: dd Month yyyy}

\def\beq{\begin{equation}}
\def\eeq{\end{equation}}
\def\bea{\begin{eqnarray}}
\def\eea{\end{eqnarray}}
\def\ba{\begin{array}}
\def\ea{\end{array}}
\def\bi{\begin{itemize}}
\def\ei{\end{itemize}}

\def\bc{\begin{center}}
\def\ec{\end{center}}
\def\bs{\begin{small}}
\def\es{\end{small}}
\def\bfs{\begin{footnotesize}}
\def\efs{\end{footnotesize}}
\def\bt{\begin{tiny}}
\def\et{\end{tiny}}

\begin{document}

\title{On the Schwarzschild-de Sitter metric of  nonlocal de Sitter gravity}

\author[affil1]{Ivan Dimitrijevic}
\ead{ivand@matf.bg.ac.rs}
\author[affil2,affil3]{Branko Dragovich}
\ead{dragovich@ipb.ac.rs}
\author[affil1]{Zoran Rakic}
\ead{zrakic@matf.bg.ac.rs}
\author[affil4]{Jelena Stankovic}
\ead{jelenagg@gmail.com}
\address[affil1]{Faculty of Mathematics, University of Belgrade, Studentski trg 16,  Belgrade, Serbia}
\address[affil2]{Institute of Physics, University of Belgrade, Pregrevica 118, 11080 Belgrade, Serbia}
\address[affil3]{Mathematical Institute, Serbian Academy of Sciences, Kneza Mihaila 36, Belgrade, Serbia}
\address[affil4]{Teacher Education Faculty, University of Belgrade, Kraljice Natalije 43, Belgrade, Serbia}
\newcommand{\AuthorNames}{Ivan Dimitrijevic et al.}

\newcommand{\FilMSC}{83D05, 53B21,  53B50; Secondary 53C25, 83F05}
\newcommand{\FilKeywords}{nonlocal gravity, equations of motion, de Sitter space, Schwarzschild-type metric, pseudo-Riemannian manifold. }
\newcommand{\FilCommunicated}{(name of the Editor, mandatory)}

\begin{abstract}
Earlier constructed a simple nonlocal de Sitter gravity model has a cosmological solution in a very good agreement with astronomical observations. In this paper, we continue the investigation of the nonlocal de Sitter model of gravity, focusing on finding an appropriate solution for the Schwarzschild-de Sitter metric. We succeeded to solve the equations of motion in a certain approximation. The obtained approximate solution is of particular interest for examining the possible role of non-local de Sitter gravity in describing the effects in galactic dynamics that are usually attributed to dark matter.
\end{abstract}

\maketitle

\makeatletter
\renewcommand\@makefnmark%
{\mbox{\textsuperscript{\normalfont\@thefnmark)}}}
\makeatother

\section{Introduction}

 For more than 100 years, General Relativity (GR) has been considered as one of the most beautiful and successful physical theories  -- from a phenomenological point of view, several remarkable  predictions have been confirmed \cite{ellis}. The Standard Model of Cosmology (SMC) assumes that GR is  valid and applicable not only in the Solar system but also at the galactic and cosmological scale. However, GR has not been verified at the galactic and cosmological scales without the use of Dark Matter (DM) and Dark Energy (DE). Furthermore, the SMC assumes that the universe contains about 68 \% of DE, 27 \% of DM and only 5 \% of ordinary matter.
   Also, GR solutions for the black holes as well as for the beginning of the universe contain singularity and it means that GR should be modified in the vicinity of these singularities. It should be also mentioned that  GR  is nonrenormalizable theory from the quantum point of view. Hence, in spite of significant successes of GR, it is reasonable to doubt in its validity in description and understanding of all  astrophysical and cosmological gravitational phenomena.  Keeping all this in mind, it follows that general relativity is not a final theory of gravitation and that its expansion should be considered.

 Since, there is still no  physical principle that could suggest us in which direction we should search for an extension of GR, there are many approaches to its modification, see \cite{faraoni,nojiri,clifton,nojiri1,capozziello,dragovich0} as some reviews. Despite many attempts, there is not yet generally accepted modification of general relativity. One of the current and attractive approaches is nonlocal modified gravity, see, e.g. \cite{biswas1,biswas3,biswas4,biswas5,deser}. In an analytic nonlocal  modification, the Einstein-Hilbert action is extended by a term that contains all higher order degrees of d'Alembert-Beltrami operator  $\Box = \nabla_\mu \nabla^\mu = \frac{1}{\sqrt{-g}} \partial_\mu (\sqrt{-g} g^{\mu\nu}\partial_\nu) .$ Usually, $\Box$ is used in the form of an analytic expression $F(\Box) = \sum_{n=0}^{+\infty} f_n \ \Box^n ,$ see \cite{koshelev2,koshelev3,eliz,koivisto,capozziello2009,dimitrijevic10}. Recently, we used nonlocal operator in the form $ F(\Box) = 1+ \mathcal{F}(\Box) = 1 + \sum_{n=1}^{+\infty} f_n \ \Box^n + \sum_{n= 1}^{+\infty} f_{-n} \ \Box^{-n}$ \cite{dimitrijevic13}. Some other typical nonlocal models can be seen in \cite{woodard,maggiore,barvinsky2012,modesto2}.

 So far we have considered the nonlocal de Sitter models with analytic nonlocality based on the action given by

\begin{align} \label{eq1.1}
S = \frac{1}{16 \pi G} \int d^4 x \ \sqrt{-g}\ \big(R- 2 \Lambda  + P(R)\ \mathcal{F}(\Box)\ Q(R)\big) ,
\end{align}
where $\Lambda$ is the cosmological constant, $P(R)$ and $Q(R)$ are some differentiable functions of the Ricci scalar $R$, see \cite{dimitrijevic1,dimitrijevic2,dimitrijevic3,dimitrijevic4,dimitrijevic5,dimitrijevic6,dimitrijevic7,dimitrijevic8,dimitrijevic9,dimitrijevic10,
dimitrijevic11,dimitrijevic12} and references therein. The case $P(R) = Q(R) =R$ and $\Lambda =0$
has attracted a lot of attention, e.g. nonlocal $R^2$ gravity, see \cite{biswas1,koshelev2} and references therein. In \cite{dimitrijevic10}
a very special case  is  $P(R) = Q(R) = \sqrt{R - 2\Lambda}$ was considered:

\begin{align} \label{eq1.2}
S = \frac{1}{16 \pi G} \int d^4 x \ \sqrt{-g}\ \big(R - 2 \Lambda + \sqrt{R - 2\Lambda}\ \mathcal{F}(\Box)\ \sqrt{R - 2\Lambda} \big),
\end{align}
where $ \mathcal{F}(\Box) = \sum_{n=1}^{+\infty} f_n \Box^n .$
Importance of  this model is not only in its simple form but also in its cosmological solutions in flat space-time: (i) $a_1(t) = A t^{\frac{2}{3}} e^{\frac{\Lambda}{14} t^2} $ and (ii) $a_2(t)= A e^{\frac{\Lambda}{6} t^2} .$
Solution $a_1(t)$  mimics interplay of dark matter and dark energy in a good agreement with  cosmological observations \cite{dimitrijevic10}.
Solution $a_2 (t)$ is the nonsingular bounce one. Nonlocal gravity model \eqref{eq1.2} also contains  several other vacuum cosmological solutions in flat, closed and open space, not only with $ \mathcal{F}(\Box) = \sum_{n=1}^{+\infty} f_n \Box^n $ but also when $\FF(\Box) =   \sum_{n=1}^{+\infty} \big( f_n \Box^n  +  f_{-n} \Box^{-n} \big), $ see \cite{dimitrijevic13}. After very successful application of model \eqref{eq1.2} with respect to dark energy and dark matter at the cosmological scale, it is natural to be interested how this model works at smaller scales.

  In this paper we investigate  model \eqref{eq1.2} with static metric around spherically symmetric body (stellar, galactic). In other words, we are interested in the Schwarzschild-type metric in nonlocal de Sitter gravity given by \eqref{eq1.2}.  The corresponding  nonlocal operator has the form
  \begin{align} \label{eq1.3}
\mathcal{F}(\Box) =   \sum_{n=1}^{+\infty} \big( f_n \Box^n  +  f_{-n} \Box^{-n} \big).
\end{align}
Here we want  to find metric solution around a massive, non-rotating, without charge, and spherically symmetric object. It is known that in GR with the cosmological constant $\Lambda$ (de Sitter gravity) such solution  is related to the Schwarzschild-de Sitter metric,
\begin{equation}\label{eq1.4}
  \mathrm ds^2 = - A(r)\mathrm dt^2 + A(r)^{-1} \mathrm dr^2 + r^2\mathrm d\theta^2 + r^2\sin^2 \theta \mathrm d\varphi^2,
\end{equation}
where $A(r)=1-\frac{2 M G}{r} - \frac{\Lambda r^2}{3}$ (the speed of light in the vacuum is taken $c=1$).

 Structure of the paper is as follows. Section \ref{Sec.2} contains an introduction to our very simple nonlocal de Sitter gravity model, $P(R) = Q(R) = \sqrt{R - 2\Lambda} .$ In Section \ref{Sec.3}, we look for the Schwarzschild-type metric and we find a solution  using the corresponding eigenvalue problem.   In Section \ref{Sec.4}, we discuss obtained  solution of this nonlocal de Sitter gravity and give some concluding remarks with our plan for future investigation. There is also an appendix which contains derivation of some useful formulas.

\section{On nonlocal de Sitter gravity model}       
\label{Sec.2}

 Let us mention some facts of our nonlocal gravity model (see \cite{dimitrijevic13}) which is given by the action

\begin{align} \label{eq2.1}
S = \frac{1}{16 \pi G} \int d^4 x \ \sqrt{-g}\ \sqrt{R - 2\Lambda}\ F(\Box)\ \sqrt{R - 2\Lambda} ,
\end{align}
where $F (\Box)$ is the following formal expansion in terms of the d'Alemberian  $\Box$:

\begin{align}  \label{eq2.2}
F(\Box) = 1+ \FF (\Box) = 1 + \FF_{+} (\Box) + \FF_{-} (\Box) , \quad \FF_{+} (\Box) =\sum_{n=1}^{+\infty} f_n \ \Box^n , \ \FF_{-} (\Box) =\sum_{n= 1}^{+\infty} f_{-n} \ \Box^{-n} .
\end{align}
When $F(\Box)= 1$, i.e. $\FF (\Box) = 0$, then  model \eqref{eq2.1} becomes
local and coincides with Einstein-Hilbert action with cosmological constant $\Lambda$:

\begin{align} \label{eq2.3}
  S_0 = \frac{1}{16 \pi G} \int d^4 x \ \sqrt{-g}\ \sqrt{R - 2\Lambda} \ \sqrt{R - 2\Lambda} = \frac{1}{16 \pi G} \int d^4 x \ \sqrt{-g}\ (R - 2 \Lambda) .
\end{align}
 It is worth pointing out that action \eqref{eq2.1} can be obtained in a very simple and natural way from action \eqref{eq2.3}
 by embedding nonlocal operator \eqref{eq2.2} in symmetric product form of $R- 2\Lambda$, that is $\sqrt{R - 2 \Lambda}\ \sqrt{R - 2 \Lambda} .$
Action \eqref{eq2.1} does not contain matter term and this is intentionally done to better see possible role of this nonlocal model  in effects
usually assigned to dark matter and dark energy.

 The  equations of motion (EoM) for model \eqref{eq1.1}, for $P(R)=Q(R)$ are given by (for more detail, see \cite{dimitrijevic9,dimitrijevic13}):

\begin{align} \label{eq2.7a}
  &  \hspace{10mm}G_{\mu\nu}+ \Lambda g_{\mu\nu} - \frac{g_{\mu\nu}}{2} P(R) \FF (\Box) P(R) + R_{\mu\nu} W - K_{\mu\nu} W + \frac 12 \Omega_{\mu\nu} = 0 , \\
& \hspace{10mm}W = 2 P'(R)\ \FF (\Box)\ P(R), \quad K_{\mu\nu} = \nabla_\mu \nabla_\nu - g_{\mu\nu}\Box , \label{eq2.7b} \\
&\hspace{10mm}\Omega_{\mu\nu} =  \sum_{n=1}^{+\infty} f_n \sum_{\ell=0}^{n-1} S_{\mu\nu}(\Box^\ell P, \Box^{n-1-\ell} P)  -\sum_{n=1}^{+\infty} f_{-n} \sum_{\ell=0}^{n-1} S_{\mu\nu}(\Box^{-(\ell+1)} P, \Box^{-(n-\ell)} P). \label{eq2.7c}
\end{align}

 If $P(R)$ is an eigenfunction of the corresponding d'Alembert-Beltrami operator $\Box$, and consequently also of its inverse $\Box^{-1}$, i.e.  holds

\begin{align} \label{eq2.8}
\Box P(R) = q  P (R), \qquad \Box^{-1} P(R) = q^{-1} P(R) ,  \qquad \FF (\Box) P(R) = \FF (q) P(R) , \qquad q\neq 0 ,
\end{align}
 then

\begin{align}
  &\hspace{10mm}W = 2 \FF(q) P' P,  \qquad \FF(q) = \sum_{n=1}^{+\infty} f_n \ q^n + \sum_{n=1}^{+\infty} f_{-n} \ q^{-n} , \label{eq2.8} \\
 &\hspace{10mm}S_{\mu\nu}(\Box^{\ell} P, \Box^{(n- 1 -\ell)} P)  = q^{n-1} S_{\mu\nu} (P, P),  \label{eq2.8a} \\
 &\hspace{10mm}S_{\mu\nu}(\Box^{-(\ell+1)} P, \Box^{-(n-\ell)} P) = q^{-n-1} S_{\mu\nu} (P, P),  \label{eq2.8b} \\
 &\hspace{10mm}S_{\mu\nu}(P, P) = g_{\mu\nu} \big(\nabla^\alpha P \ \nabla_\alpha P + P \Box P \big) - 2\nabla_\mu P\ \nabla_\nu P,  \label{eq2.8b}\\
 &\hspace{10mm}\Omega_{\mu\nu} = \FF'(q) S_{\mu\nu}(P,P), \qquad \FF'(q) = \sum_{n=1}^{+\infty} n\ f_n \ q^{n-1} -
 \sum_{n=1}^{+\infty} n\ f_{-n} \ q^{-n-1} , \label{eq2.9}
\end{align}
and
\begin{equation} \label{eq2.10}
  G_{\mu\nu}+ \Lambda g_{\mu\nu} - \frac{g_{\mu\nu}}{2}  \FF (q)P^2 + 2 \FF(q) R_{\mu\nu} P P' - 2 \FF(q) K_{\mu\nu} P P' + \frac 12 \FF'(q) S_{\mu\nu}(P,P) = 0.
\end{equation}
The last equation becomes
\bea  \hspace{-7mm}
  \left(G_{\mu\nu}+ \Lambda g_{\mu\nu}\right)\left(1 + 2 \FF(q) P P'\right) + \FF(q)g_{\mu\nu}\left(-\frac 12 P^2 + P P'(R-2\Lambda)\right) - 2 \FF(q) K_{\mu\nu} P P' + \frac 12 \FF'(q) S_{\mu\nu}(P,P) = 0. \label{eq2.11}
\eea

 If $P(R)= \sqrt{R-2\Lambda}$, then  $P(R)P'(R) = \frac 12 $ and

\begin{equation}
\Box \sqrt{R - 2\Lambda} = q\,  \sqrt{R - 2\Lambda} = \eta\, \Lambda  \sqrt{R - 2\Lambda} , \qquad \eta\,\Lambda \neq 0 , \label{eq2.11a}
\end{equation}
where $q = \eta\, \Lambda$ and $q^{-1}= \eta^{-1} \Lambda^{-1}$ \ ($\eta$ -- dimensionless) follows from dimensionality of equalities \eqref{eq2.11a}.
Since $P(R)= \sqrt{R-2\Lambda}$,  EoM \eqref{eq2.11} simplify to

\begin{equation}  \label{eq2.12}
  \left(G_{\mu\nu}+ \Lambda g_{\mu\nu}\right)\left(1 + \FF(q)\right) + \frac 12 \FF'(q) S_{\mu\nu}(\sqrt{R-2\Lambda},\sqrt{R-2\Lambda}) = 0.
\end{equation}

 It is clear that EoM \eqref{eq2.12} are satisfied if

\begin{align} \label{eq2.13}
\mathcal{F} (q) = -1  \qquad \qquad\text{and} \qquad \qquad  \mathcal{F}' (q) = 0 .
\end{align}

It is worth pointing out that not only nonlocal de Sitter  model \eqref{eq2.1} is very simple and natural but also such are corresponding EoM \eqref{eq2.12}  with respect to all other models and their EoM that can be derived from  \eqref{eq1.1} with $\Lambda \neq 0$.

 Let us remark that nonlocal operator $\mathcal{F} (\Box)$, which satisfies conditions \eqref{eq2.13} in model \eqref{eq2.1},
can be taken in the symmetric form

\begin{equation}
F(\Box) = 1 +  \mathcal{F} (\Box), \qquad \mbox{where} \qquad\mathcal{F} (\Box) = \sum_{n=1}^{+\infty} \tilde{f}_n \Big[\Big(\frac{\Box}{q}\Big)^n  + \Big(\frac{q}{\Box}\Big)^n \Big] , \label{eq2.13a}
\end{equation}
where $\tilde{f}_n$ are dimensionless coefficients.
It is easy to prove that
 $\mathcal{F}(\Box)$  presented in the following symmetric form:

\begin{align} \label{eq2.14}
 \mathcal{F}(\Box) = \sum_{n=1}^{+\infty} \tilde{f}_n \Big[\Big(\frac{\Box}{q}\Big)^n  + \Big(\frac{q}{\Box}\Big)^n \Big] = - \frac{1}{2 e}\, \Big(\frac{\Box}{q} e^{\frac{\Box}{q}} + \frac{q}{\Box} e^{\frac{q}{\Box}}\Big) , \qquad q \neq 0 ,
\end{align}
satisfies conditions \eqref{eq2.13}.
We can take  the eigenvalue $q$ of d'Alembertian $\Box$
  to be proportional to $\Lambda$, i.e. $q = \eta\, \Lambda$, where $\eta\, \neq 0$ (since $\Lambda \neq 0$) is a definite dimensionless constant. Moreover,
it has to be $q = \eta\, \Lambda$, since there is no other parameter than $\Lambda$ which is of the same  dimension as $\Box$  in this nonlocal
gravity model. Hence, nonlocal operator \eqref{eq2.14} can be rewritten as

\begin{align} \label{eq2.16}
 \mathcal{F} (\Box) = \sum_{n=1}^{+\infty} \tilde{f}_n \Big[\Big(\frac{\Box}{\eta\Lambda}\Big)^n  + \Big(\frac{\eta\Lambda}{\Box}\Big)^n \Big] = - \frac{1}{2 e} \Big(\frac{\Box}{\eta \Lambda} e^{\frac{\Box}{\eta \Lambda}} + \frac{\eta \Lambda}{\Box} e^{\frac{\eta \Lambda}{\Box}}\Big) , \qquad \eta\,\Lambda \neq 0,
 \end{align}
where for some specific $\,\Box\,$ holds  $\,\Box \sqrt{R - 2\Lambda} = \eta\, \Lambda \sqrt{R - 2\Lambda} .$

\section{Schwarzschild-de Sitter-type metric}
\label{Sec.3}

 We want to investigate our model outside the spherically symmetric massive body. Since this model is a nonlocal generalization of general relativity with the cosmological constant $\Lambda$, it is natural to consider a generalization of the Schwarzschild-de Sitter (SdS) metric starting from the standard Schwarzschild expression

\begin{equation} \label{eq3.1}
  \mathrm ds^2 = - A(r)\mathrm dt^2 + B(r) \mathrm dr^2 + r^2\mathrm d\theta^2 + r^2\sin^2 \theta \mathrm d\varphi^2.
\end{equation}

The corresponding scalar curvature $R$ of above metric \eqref{eq3.1} is

\bea \label{eq3.2} R= \frac{2}{r^2}-\frac{2}{r^2 B(r)}-\frac{2 A'(r)}{r A(r)B(r)}+\frac{A'(r)^2}{2 A(r)^2 B(r)}+\frac{2 B'(r)}{r B^2(r)}+\frac{A'(r)B'(r)}{2 A(r)B(r)^2}-\frac{A''(r)}{A(r)B(r)}\,\eea
where ' denotes derivative with respect to $r.$\vspace{1.5mm}
 In this paper we will consider the case $B(r)= A(r)^{-1}$, and
 formula \eqref{eq3.2} becomes

 \bea \label{eq3.3} R = \frac{2-2A(r)-4 r A'(r)-r^{2}A''(r)}{r^{2}}.\eea
 Note that \eqref{eq3.3} can be rewritten in the more  compact form

 \bea \label{eq3.3a} R(r) =\frac{1}{r^2} \frac{\partial^2}{\partial r^2}\big[r^2 \big(1-A(r) \big)\big]. \eea

 According to Section \ref{Sec.2}, to find a solution of equations of motion it is necessary to solve an eigenvalue problem \eqref{eq2.11a}, that is $\Box \sqrt{R - 2\Lambda} = q\,  \sqrt{R - 2\Lambda}$. Note that here d'Alembertian $\Box$ acts in the following way:

 \bea\label{eq3.4} \Box u(r) = A(r)\,u''(r)+ (A'(r) + \frac{2}{r}\;A(r))\,u'(r)= \frac{1}{r^2} \frac{\partial}{\partial r}\big[r^2 A(r) \frac{\partial u}{\partial r}  \big] ,\eea
where $u(r)$ is any differentiable scalar function.

 Let us now consider function $A(r)$ in the  form

 \bea\label{eq3.4a}  A(r) = 1 -\frac{\mu}{r} -\frac{\nu}{r^2} - \frac{\Lambda}{3} r^2 - f(r),\eea where $\mu$ and $\nu$ are some parameters to be discussed later. Then one can show that for $A(r)$ given by
 \eqref{eq3.4a}  holds

\bea \label{eq3.5} R(r) =\frac{1}{r^2} \frac{\partial^2}{\partial r^2}\big[r^2 \big(1-A(r)\big)\big]  = 4 \Lambda + \frac{1}{r^2} \frac{\partial^2}{\partial r^2}\big[r^2 f(r)\big]. \eea

 Denoting  $u(r) = \sqrt{R- 2\Lambda} $ and using expression \eqref{eq3.4} for d'Alembertian,  the corresponding equation $\Box \sqrt{R - 2\Lambda} = q\,  \sqrt{R - 2\Lambda}$ becomes

\bea   \label{eq3.6} \Box \sqrt{R- 2\Lambda}  = \frac{1}{r^2} \frac{\partial}{\partial r}\big[r^2 A(r) \frac{\partial }{\partial r} \sqrt{R- 2\Lambda}  \big] = q\, \sqrt{R- 2\Lambda}. \eea
Since unknown function $A(r)$ is contained also in the scalar curvature $R(r)$, equation \eqref{eq3.6} is very complicated and difficult to find exact solution. To get an approximative  solution we take $A(r) \thickapprox 1$ in \eqref{eq3.6}, what is applicable when

\begin{align}
\big|\frac \mu r \big|\ll 1, \quad \big|\frac \nu{r^{2}}\big|\ll 1, \quad |\Lambda r^2|\ll 1, \quad  |f(r)|\ll 1,   \label{eq3.7}
\end{align}
in units $c=1$. Under conditions \eqref{eq3.7}, equation \eqref{eq3.6} becomes

\bea    \label{eq3.8} \triangle \sqrt{R- 2\Lambda}  = \frac{1}{r^2} \frac{\partial}{\partial r}\big[r^2  \frac{\partial }{\partial r} \sqrt{R- 2\Lambda}  \big] = q\, \sqrt{R- 2\Lambda} , \eea
where

\begin{align}  \label{eq3.9} \triangle   = \frac{1}{r^2} \frac{\partial}{\partial r}\big[r^2  \frac{\partial  }{\partial r}  \big] = \frac{\partial^2}{\partial r^2} + \frac{2}{r} \frac{\partial}{\partial r}
\end{align}
is the Laplace operator (Laplacian) in spherical coordinates.

One can easily check that the general solution of equation

\begin{align} \label{eq3.10}
\triangle u(r) = \frac{\partial^2 u(r)}{\partial r^2} + \frac{2}{r} \frac{\partial u(r)}{\partial r} =q\ u(r)
\end{align}
is

\begin{align}
 \label{eq3.11}   u(r) = \frac{C_1}{r}\ e^{\sqrt{q}\ r} + \frac{C_2}{r}\ e^{-\sqrt{q}\ r} \, .
\end{align}
Since the metric should tend to the Minkowski one at large distances, in the sequel we will use only solution

\begin{align}
 \label{eq3.12}   u(r) = \frac{C_2}{r}\ e^{-\sqrt{q}\ r} .
\end{align}

According to equation \eqref{eq3.5} we have

\bea \label{eq3.13} R(r) -2 \Lambda = 2\Lambda + \frac{1}{r^2} \frac{\partial^2}{\partial r^2}\big[r^2 f(r)\big] = u^2(r). \eea
Equation \eqref{eq3.13} can be rewritten in the  form

\bea   \label{eq3.14} r^2 f^{''}(r) + 4 r f'(r) +2 f(r) = -2\Lambda r^2 + C_2^2\ e^{-2\sqrt{q}\ r}. \eea
General solution of equation \eqref{eq3.14} is

\begin{align}
\label{eq3.15} f(r) = -\frac{\Lambda}{6} r^2 + \frac{C_2^2}{4 q}\ \frac{1}{r^2}\ e^{-2\sqrt{q}\ r} + \frac{C_3}{r} + \frac{C_4}{r^2} .
\end{align}

Replacing $f(r)$ in \eqref{eq3.4a} by expression \eqref{eq3.15} one obtains

\bea\label{eq3.16}  A(r) = 1 -\frac{\mu}{r} -\frac{\nu}{r^2} - \frac{\Lambda r^2}{6}  -  \frac{C_2^2}{4 q r^2}\ e^{-2\sqrt{q}\ r} - \frac{C_3}{r} - \frac{C_4}{r^2} .\eea
Since all parameters $\mu, \nu, C_3, C_4$ have been so far arbitrary we can take $C_3 = C_4 = 0$ and maintain only $\mu$ and $\nu$, i. e. in the sequel we have

\bea\label{eq3.17}  A(r) = 1 -\frac{\mu}{r} -\frac{\nu}{r^2} - \frac{\Lambda r^2}{6} -  \frac{C_2^2}{4 q r^2} \ e^{-2\sqrt{q}\ r} .\eea
Then the corresponding scalar curvature becomes

\begin{align} \label{eq3.18}
 R(r) = 2 \Lambda + \frac{C_2^2}{r^2}\ e^{-2\sqrt{q}\ r} .
\end{align}

\section{Discussion and conclusion}
\label{Sec.4}

 In the previous sections we considered  nonlocal de Sitter gravity model defined by its action \eqref{eq2.1} at scales smaller than the cosmological one, i.e. related to stars, galaxies, and
clusters of galaxies. Cosmological solutions of model \eqref{eq2.1} are presented in \cite{dimitrijevic10,dimitrijevic13}, and it was shown that the cosmological solution in flat space-time, $a_1(t) = A t^{\frac{2}{3}} e^{\frac{\Lambda}{14} t^2} $ gives good agreement with observational data. Consequently, it is natural to see how this nonlocal model works at the  smaller scales.

It is well known that $A (r)$ of the standard Schwarzschild-de Sitter metric, i.e. in the case of  local de Sitter gravity, is

\begin{align}
\label{eq4.1}   A_{\ell} (r)  = 1 - \frac{2 G M}{c^2 r} - \frac{\Lambda r^2}{3 c^2}  , \qquad r \geq r_0 ,
\end{align}
where $r_0$ is the radius of spherically symmetric massive body ($M$--mass), and $G$ is the Newton gravitational  constant.  Note that \eqref{eq4.1} is written in the international system of units (SI). The nonlocal version of $A(r)$ \eqref{eq3.16} can be rewritten as

\bea\label{eq4.2}  A_{n\ell}(r) = 1 - \frac{2 G M}{c^2 r} - \frac{\Lambda r^2}{6 c^2}    + \frac{\delta^2}{q r^2}\big(1 -   e^{-2\sqrt{q}\ r}\big) , \qquad  q = \frac{\eta \Lambda}{c^2} ,\eea
where we joined terms $\frac{\nu}{r^2}$ and $\frac{C_2^2}{4 q r^2}\ e^{-2\sqrt{q}\ r}$ taking $ \delta^2 = \frac{C_2^2}{4}, \nu = -\frac{\delta^2}{ q}$.    It is worth noting that $\delta$ and $\eta$ are dimensionless parameters, and their values should be determined by experiments (astronomical observations). Comparing \eqref{eq4.1} and \eqref{eq4.2} follows formula

\begin{align} \label{eq4.3}
A_{n\ell}(r) - A_{\ell}(r) =  \frac{\Lambda r^2}{6 c^2}    + \frac{\delta^2}{q r^2}\big(1 -   e^{-2\sqrt{q}\ r}\big) .
\end{align}

Testing  formula \eqref{eq4.2} in the Solar and other astronomical systems  is one of our main next tasks, and the results will be presented elsewhere.

At the end, it is worth noting that in paper \cite{dragovich2} was found connection between cosmological constant $\Lambda$ and mass of a scalar particle which has its origin in $p$-adic string theory \cite{dragovich1}. About cosmological research with application of nonlocal fields in the matter sector of general relativity one can see \cite{arefeva2,koshelev2011} and references therein.

\section*{Appendix}
In the appendix we present curvature tensors for the Schwarzschild-de Sitter-type metric \eqref{eq3.1}

\begin{equation}
  \mathrm ds^2 = - A(r)\mathrm d t^2 + B(r)\mathrm d r^2 + r^2 \mathrm d \theta^2 + r^2 \sin^2 \theta \mathrm d\varphi^2.
\end{equation}

The Christoffel symbols are given as follows (unlisted ones are equal to zero and $'$ denotes derivative with respect to $r$):
\begin{equation}
  \begin{aligned}
    \Gamma_{01}^0 & = \frac 12 \frac{A'}A, & \Gamma_{00}^1 & = \frac 12 \frac{A'}B, & \Gamma_{11}^1 & = \frac 12 \frac{B'}B, & \Gamma_{22}^1 & = - \frac rB, & \Gamma_{33}^1 & = - \frac{r\sin^2\theta}B,\\
    \Gamma_{12}^2 & = \frac 1r, & \Gamma_{33}^2 &= - \sin\theta \cos\theta, & \Gamma_{13}^3 &= \frac 1r, &\Gamma_{23}^3 &= \cot \theta .
  \end{aligned}
\end{equation}

Curvature tensor components with three or four different indices are all equal to zero. Only nonzero components are with two different indices and they are listed below:
\begin{equation}
  \begin{aligned}
    R_{0101} & = \frac A4\left(- \left(\frac{A'}A\right)^2 -\frac{A'}A \frac{B'}B + 2\frac{A''}A\right), &
    R_{0202} & = \frac r2 \frac{A'}B, &
    R_{0303} & = \frac r2 \frac{A'}B \sin^2 \theta, \\
    R_{1212} & = \frac r2 \frac{B'}B, &
    R_{1313} & = \frac r2 \frac{B'}B \sin^2 \theta, &
    R_{2323} & = r^2 \frac{B-1}B \sin^2 \theta. \\
  \end{aligned}
\end{equation}

The Ricci tensor is diagonal and its components are:
\begin{equation}
  \begin{aligned}
    R_{00}&= \frac{A''}{2 B} - \frac{A'B'}{4 B^2}-\frac{A'^2}{4 AB}+\frac{A'}{r B}, &
    R_{11}&= -\frac{A''}{2 A}+ \frac{A'B'}{4 A(r) B(r)} + \frac{A'^2}{4 A^2} + \frac{B'}{r}, \\
    R_{22}&= -\frac{r A'}{2 AB}+\frac{r B'}{2 B^2}-\frac{1}{B}+1, &
    R_{33}&= \left( -\frac{r A'}{2 AB}+\frac{r B'}{2 B^2}-\frac{1}{B}+1\right)\sin^2 \theta.
  \end{aligned}
\end{equation}

The scalar curvature is
\begin{equation}
  R = -\frac{A''}{AB}+\frac{A'B'}{2 A B^2}+\frac{A'^2}{2 A^2 B}-\frac{2 A'}{r AB}+\frac{2 B'}{r B^2}-\frac{2}{r^2 B}+\frac{2}{r^2} .
\end{equation}

The Einstein tensor is diagonal and components are presented as

\begin{equation}
  \begin{aligned}
    G_{00}&= \frac{A B'}{r B^2}-\frac{A}{r^2 B} + \frac{A}{r^2}, &
    G_{11}&= \frac{A'}{r A}-\frac{B}{r^2}+\frac{1}{r^2}, \\
    G_{22}&= \frac{r^2 A''}{2 AB}-\frac{r^2 A' B'}{4 A B^2}-\frac{r^2 A'^2}{4 A^2 B}+\frac{r A'}{2 AB}-\frac{r B'}{2 B^2}, &
    G_{33}&= \left(\frac{r^2 A''}{2 AB}-\frac{r^2 A' B'}{4 A B^2}-\frac{r^2 A'^2}{4 A^2 B}+\frac{r A'}{2 AB}-\frac{r B'}{2 B^2}\right)\sin^2 \theta.
  \end{aligned}
\end{equation}

\subsection*{{\bf Case $B = \frac 1A$}.}

In particular, for $B= \frac 1A$ we have
\begin{equation}
  \mathrm ds^2 = - A(r)\mathrm d t^2 + \frac 1{A(r)}\mathrm d r^2 + r^2 \mathrm d \theta^2 + r^2 \sin^2 \theta \mathrm d\varphi^2.
\end{equation}

The Christoffel symbols are:
\begin{equation}
  \begin{aligned}
    \Gamma_{01}^0 & = \frac 12 \frac{A'}A, &
    \Gamma_{00}^1 & = \frac 12 A A', & \Gamma_{11}^1 & = -\frac 12 \frac{A'}A, &
    \Gamma_{22}^1 & = - rA, & \Gamma_{33}^1 & = - r A\sin^2 \theta, \\
    \Gamma_{12}^2 & = \frac 1r, & \Gamma_{33}^2 &= - \sin\theta \cos\theta, &
    \Gamma_{13}^3 &= \frac 1r, &\Gamma_{23}^3 &= \cot \theta .
  \end{aligned}
\end{equation}

Curvature tensor components are:
\begin{equation}
  \begin{aligned}
    R_{0101} & = \frac 12A'', &
    R_{0202} & = \frac r2 AA', &
    R_{0303} & = \frac r2 AA' \sin^2 \theta, \\
    R_{1212} & = -\frac r2 \frac{A'}A, &
    R_{1313} & = -\frac r2 \frac{A'}A \sin^2 \theta, &
    R_{2323} & = r^2 (1-A) \sin^2 \theta. \\
  \end{aligned}
\end{equation}

The Ricci tensor is diagonal and its components are:
\begin{equation}
  \begin{aligned}
    R_{00}&= \frac{1}{2} A A''+ \frac 1r AA', &
    R_{11}&= -\frac 12 \frac{A''}{A}- \frac 1r \frac{A'}A, &
    R_{22}&= 1 - A - r A', &
    R_{33}&= \left(1 - A - r A'\right)\sin^2 \theta.
  \end{aligned}
\end{equation}

The scalar curvature is
\begin{equation}
  R = - A'' - \frac 4r A' - \frac 2{r^2}A + \frac 2{r^2}.
\end{equation}

The Einstein tensor is presented as follows:

\begin{equation}
  \begin{aligned}
    G_{00}&= -\frac{A(r) A'(r)}{r}-\frac{A(r)^2}{r^2}+\frac{A(r)}{r^2}, &
    G_{11}&= \frac{A'(r)}{r A(r)}-\frac{1}{r^2 A(r)}+\frac{1}{r^2}, \\
    G_{22}&= \frac{1}{2} r^2 A''(r)+r A'(r), &
    G_{33}&= \left( \frac{1}{2} r^2 A''(r)+r A'(r) \right)\sin^2 \theta.
  \end{aligned}
\end{equation}

\section*{Acknowledgement}

 This research was partially funded by the Ministry of Education, Science and Technological
Developments of the Republic of Serbia: grant number 451-03-68/2022-14/200104 with
Faculty of Mathematics, and grant number 451-03-1/2022-14/4 with Teacher Education
Faculty, University of Belgrade. It is also partially supported by the COST Action: CA21136 -- Addressing observational tensions in cosmology with systematics and fundamental physics (CosmoVerse).


\begin{thebibliography}{93}


\bibitem{ellis} G. F. R. Ellis, \emph{100 years of general relativity}, (2015) [arXiv:1509.01772 [gr-qc]].



\bibitem{faraoni} T. P. Sotiriou and V. Faraoni, \emph{$f(R)$ theories of gravity}, Rev. Mod. Phys. 82 (2010) 451 [arXiv:0805.1726v4 [gr-qc]].

\bibitem{nojiri}  S. Nojiri and S. D.  Odintsov,  \emph{Unified cosmic history in modified gravity: from $F(R)$ theory
to Lorentz non-invariant models}, Phys. Rep. 505 (2011) 59--144 [arXiv:1011.0544v4 [gr-qc]].

\bibitem{clifton} T. Clifton, P. G. Ferreira,  A. Padilla and  C. Skordis,  \emph{Modified gravity and cosmology},
Phys. Rep. 513 (2012) 1 [arXiv:1106.2476v2 [astro-ph.CO]].

\bibitem{nojiri1} S. Nojiri, S. D.  Odintsov and V. K. Oikonomou,  \emph{Modified gravity theories on a nutshell: Inflation, bounce and late-time evolution}, Phys. Rep. 692 (2017) 1--104 [arXiv:1705.11098 [gr-qc]].

\bibitem{capozziello}  S. Capozziello and F. Bajardi, \emph{Nonlocal gravity cosmology: An overview}, Int. J. Mod. Phys. D  31  (2022) 2230009	[arXiv:2201.04512 [gr-qc]].

\bibitem{dragovich0} B. Dragovich,  \emph{On Nonlocal modified gravity and cosmology}, Springer Proc. Mathematics
$\&$ Statistics 111 (2014) 251–-262.

\bibitem{biswas1} T. Biswas, A.  Mazumdar and W.  Siegel,   \emph{Bouncing universes in string-inspired gravity}, JCAP 0603 (2006) 009  [arXiv:hep-th/0508194].


\bibitem{biswas3} T. Biswas,  E. Gerwick, T. Koivisto and A. Mazumdar,  \emph{Towards singularity and ghost free theories
of gravity}, Phys. Rev. Lett. 108 (2012) 031101 [arXiv:1110.5249v2 [gr-qc]].

\bibitem{biswas4} T. Biswas, A. Conroy, A. S. Koshelev and A. Mazumdar, \emph{Generalized gost-free quadratic curvature gravity}, Class. Quantum Grav. 31 (2014) 015022 [arXiv:1308.2319 [hep-th]].

\bibitem{biswas5} T. Biswas, A. S. Koshelev, A. Mazumdar and S. Yu. Vernov,  \emph{Stable bounce and inflation
in non-local higher derivative cosmology}, JCAP 08 (2012) 024  [arXiv:1206.6374v2 [astro-ph.CO]].

\bibitem{deser} S. Deser and R. Woodard, \emph{Nonlocal cosmology}, Phys. Rev. Lett. 99 (2007) 111301
	[arXiv:0706.2151 [astro-ph]].


\bibitem{koshelev2}  A. S. Koshelev, L. Modesto, L. Rachwal and A. A. Starobinsky, \emph{Occurrence of exact $R^2$ inflation in non-local UV-complete gravity}, JHEP 2016 (2016) 67  [arXiv:1604.03127 [hep-th]].

\bibitem{koshelev3} L.~Buoninfante, A.~S.~Koshelev, G.~Lambiase and A.~Mazumdar,
  \emph{Classical properties of non-local, ghost- and singularity-free gravity}, JCAP 09 (2018) 034  [arXiv:1802.00399 [gr-qc]].


\bibitem{eliz} E. Elizalde, E. O. Pozdeeva and  S. Yu. Vernov, \emph{Stability of de Sitter solutions in non-local cosmological models},
PoS QFTHEP2011 138 (2012)  [arXiv:1202.0178].

\bibitem{koivisto} A. Conroy, T. Koivisto, A. Mazumdar and A. Teimouri, \emph{Generalised quadratic curvature, non-local infrared modifications of gravity and Newtonian potentials}, Class. Quantum Grav.  32  (2015) 015024  [arXiv:1406.4998v3 [hep-th]].

\bibitem{capozziello2009} S. Capozziello,  E. Elizalde,  Sh. Nojiri and S. D. Odintsov,
\emph{Accelerating cosmologies from non-local higher-derivative gravity}, Phys. Lett. B. 671 (1)  (2009) 193-198.


\bibitem{woodard} R. P. Woodard,  \emph{Nonlocal models of cosmic acceleration}, Found. Phys. 44 (2014) 213--233 	
[arXiv:1401.0254 [astro-ph.CO]].

\bibitem{maggiore}  E. Belgacem, Y. Dirian, S. Foffa and M. Maggiore,
\emph{Nonlocal gravity. Conceptual aspects and cosmological predictions}, JCAP 03 (2018) 002
  [arXiv:1712.07066 [hep-th]].

\bibitem{barvinsky2012} A. O. Barvinsky,   \emph{Dark energy and dark matter from nonlocal ghost-free gravity theory}, Phys. Lett. B  710 (2012) 12.




  (1977) 953.


 \bibitem{modesto2}   L. Modesto and L. Rachwal,  \emph{Super-renormalizable and finite gravitational theories},
 Nucl. Phys. B 889 (2014) 228 [arXiv:1407.8036 [hep-th]].


\bibitem{dimitrijevic1}  I. Dimitrijevic, B. Dragovich, J. Grujic and Z. Rakic, \emph{On modified gravity}, Springer Proc. Mathematics $\&$  Statistics 36 (2013) 251--259 [arXiv:1202.2352 [hep-th]].

\bibitem{dimitrijevic2} I. Dimitrijevic, B. Dragovich, J. Grujic and Z. Rakic, \emph{New cosmological solutions in nonlocal modified gravity}, Rom. J. Phys. 58 (5-6) (2013) 550--559  [arXiv:1302.2794 [gr-qc]].

\bibitem{dimitrijevic3} I.~Dimitrijevic, B.~Dragovich, J.~Grujic and Z.~Rakic,  \emph{A new model of nonlocal modified gravity}, Publications de l'Institut Mathematique 94 (108) (2013) 187--196.

\bibitem{dimitrijevic4} I.~Dimitrijevic, B.~Dragovich, J.~Grujic and Z.~Rakic, \emph{Some power-law
cosmological solutions in nonlocal modified gravity}, Springer Proc. Mathematics $\&$ Statistics 111 (2014) 241--250.

\bibitem{dimitrijevic5} I. Dimitrijevic, B. Dragovich, J. Grujic, A. S. Koshelev and Z. Rakic,  \emph{Cosmology of non-local $f(R)$ gravity},	
Filomat 33 (2019) 1163 [arXiv:1509.04254v2 [hep-th]].

\bibitem{dimitrijevic6} I. Dimitrijevic, B. Dragovich, J. Stankovic, A. S. Koshelev and Z. Rakic, \emph{On nonlocal modified gravity and its cosmological solutions}, Springer Proc. Mathematics $\&$ Statistics 191 (2016) 35--51 [arXiv:1701.02090 [hep-th]].

\bibitem{dimitrijevic7} I. Dimitrijevic, B. Dragovich, J. Grujic  and Z. Rakic, \emph{Some cosmological solutions of a nonlocal modified gravity},  Filomat 29 (2015) 619  [arXiv:1508.05583 [hep-th]].

\bibitem{dimitrijevic8} I. Dimitrijevic, \emph{Cosmological solutions in modified gravity with monomial nonlocality}, Appl. Math. Comput. 285 (2016) 195.

\bibitem{dimitrijevic9} I. Dimitrijevic, B. Dragovich, Z. Rakic and J. Stankovic, \emph{Variations of infinite derivative modified gravity},  Springer Proc. Mathematics $\&$ Statistics 263  (2018)  91.

\bibitem{dimitrijevic10} I. Dimitrijevic, B.  Dragovich, A. S.  Koshelev, Z. Rakic and J. Stankovic, \emph{Cosmological solutions of a nonlocal square-root gravity}, Phys. Lett. B 797 (2019) 134848 [arXiv:1906.07560 [gr-qc]].

\bibitem{dimitrijevic11} I. Dimitrijevic, B.  Dragovich, A. S.  Koshelev, Z. Rakic and J. Stankovic,  \emph{Some cosmological solutions of a new nonlocal gravity model}, Symmetry 2020 12  (2020) 917 [arXiv:2006.16041 [gr-qc]].

\bibitem{dimitrijevic12} I. Dimitrijevic, B.  Dragovich,  Z. Rakic and J. Stankovic,  \emph{New cosmological solutions of a nonlocal gravity model}, Symmetry 2022 14 (2022) 3 [arXiv:2112.06312 [gr-qc]].

\bibitem{dimitrijevic13} I. Dimitrijevic, B.  Dragovich,  Z. Rakic and J. Stankovic,  \emph{Nonlocal de Sitter gravity and its exact cosmological solutions}, JHEP 12 (2022) 054 [arXiv:2206.13515v1 [gr-qc]].

\bibitem{dragovich2} B. Dragovich, \emph{A $p$-Adic matter in a closed universe}, Symmetry 2022  14  (2022) 73 [arXiv:2201.02200 [hep-th]].

\bibitem{dragovich1} B. Dragovich, A. Yu. Khrennikov, S. V. Kozyrev, I. V. Volovich and E. I. Zelenov,
\emph{$p$-Adic mathematical physics: the first 30 years}, $p$-Adic Numb. Ultrametric Anal. Appl.  9 (2) (2017) 87--121 [arXiv:1705.04758 [math-ph]].


\bibitem{arefeva2} I. Ya. Aref'eva and  I. V. Volovich, \emph{Cosmological daemon}, JHEP  2011 (2011) 102  [arXiv:1103.0273v2 [hep-th]].

\bibitem{koshelev2011}  A. Koshelev and S. Yu. Vernov,  \emph{Analysis of scalar perturbations in cosmological models with a non-local scalar field}, Class. Quantum Grav.  28 (2011) 085019.







  \end{thebibliography}
\end{document}